\documentclass{aa}
\usepackage{psfig}
\usepackage{graphics}

\usepackage{times}
\usepackage{mathptm}
\fontfamily{ptmcm}\selectfont

\def\gradip{\hbox{\rlap{\hbox{.}}\raise 5.truept \hbox{{\small $\circ$}}}}

\newcommand{\cdo}{$^{12}C(\alpha ,\gamma)^{16}O\;$}
\newcommand{\vhbb}{$\Delta V_{\rm HB}^{\rm Bump}\,$}

\catcode`\@=11
\def\gsim{\ifmmode{\mathrel{\mathpalette\@versim>}}
    \else{$\mathrel{\mathpalette\@versim>}$}\fi}
\def\lsim{\ifmmode{\mathrel{\mathpalette\@versim<}}
    \else{$\mathrel{\mathpalette\@versim<}$}\fi}
\def\@versim#1#2{\lower 2.9truept \vbox{\baselineskip 0pt \lineskip
    0.5truept \ialign{$\m@th#1\hfil##\hfil$\crcr#2\crcr\sim\crcr}}}
\catcode`\@=12

\begin{document}

\thesaurus{08.05.1
           08.05.3; 
           08.08.1; 
           08.12.3; 
           08.16.3; 
           10.07.2}

%\title{Observational Constraints on the Red Giant Branch Luminosity
%	Functions of Galactic Globular Clusters 
\title{Comparison between Observed and Theoretical Red Giant Branch
	Luminosity Functions of Galactic Globular Clusters 
\thanks{Based on observations with the NASA/ESA {\it Hubble Space
	Telescope}, obtained at the Space Telescope Science Institute,
	which is operated by AURA, Inc., under NASA contract
	NAS5-26555, and on observations retrieved from the ESO ST-ECF 
	Archive.}}

\author{M. Zoccali \inst{1} \and G. Piotto \inst{1}}

\offprints{M. Zoccali, e-mail: zoccali@pd.astro.it}

\institute{Dipartimento di Astronomia -- Universit\`a di Padova, 
        Padova, ITALY -- zoccali@pd.astro.it, piotto@pd.astro.it}

\date{Received ; accepted }
\titlerunning{Globular Cluster RGB luminosity functions}
\maketitle
\markboth{Zoccali and Piotto}{GGC RGB LFs}

%_________________________ ABSTRACT _____________________________________ 

\begin{abstract}

$V-$band luminosity functions (LF) have been obtained for the upper
main-sequence (MS), sub-giant branch (SGB) and red giant branch (RGB)
of 18 galactic globular clusters (GGC) from HST data. A comparison with
four sets of theoretical models has been performed. In contrast
with what was found in several previous works, a good general agreement has
been found between the observed and theoretical LF at any metallicity
[M/H]$<-0.7$. Possible discrepancies at higher metallicity, in the
upper part of the RGB, need to be confirmed with further observational
data and by extending all the models to the most metal rich regime.
The SGB shape has been used to set an upper limit to the cluster age,
and consequently a lower limit on the cluster distance. A discussion
on the still open problem of the mismatch between the observed and
theoretical RGB bump location is also presented.


\end{abstract}

%_________________________ INTRODUZIONE _____________________________________ 

\section{Introduction} \label{intro}

The luminosity function (LF) of a star cluster above the main sequence
(MS) turnoff reflects primarily the rate of luminosity evolution. This
happens because above the turnoff there is a rapid increase in the
rate at which a star luminosity changes with time, and therefore these
stars may be regarded as having started out with the same initial mass
to within a few percent.  The details of the interpretation of the
evolved star LF in terms of the relevant stellar evolution theory,
together with the many advantages of using the LFs to constrain the 
stellar models, have been excellently reviewed in a paper by Renzini 
\& Fusi Pecci (1988).

In the present paper, we will focus our attention on the LF of the red
giant branch (RGB), subgiant branch (SGB), and MS, excluding the
horizontal branch (HB) and asymptotic giant branch (AGB) stars, as
well as the blue stragglers.  Since the notation in the literature is
not homogeneous, we point out that we will call SGB the region of the
color magnitude diagram (CMD)
that goes from the MS turnoff to just after the bend where the CMD
starts to rise vertically. The RGB goes from this point up to
the brightest stars. The LF of the RGB is a simple straight line on a
magnitude-log(counts) plane (cf. Fig.~\ref{lf_pos}).  The {\it slope}
of this line allows a fundamental test of the evolutionary clock,
i.e., of the rate at which stars ascend the RGB, and it is virtually
independent of the cluster age or metallicity.  The RGB LF also
exhibits a characteristic bump, due to the hydrogen burning shell
approaching the composition discontinuity left by the deepest
penetration of the convective envelope. The level of agreement between
the empirical and theoretical determination of the magnitude of the
RGB bump, with respect to the HB level, for all the clusters of our
database, has been analyzed in a dedicated paper (Zoccali et
al. 1999), but it will be also discussed in the last section of this
paper.

The SGB is certainly the most interesting part of the evolved-star LF,
being the most sensitive to the stellar physical parameters. Due to
the rapid increase in the number of stars towards the turnoff, and to
the fact that the SGB is almost horizontal in the CMD, the SGB LF is
mainly characterized by a steep drop (in the following called the SGB
{\it break}) and by a peak just above the break (see Fig.~\ref{lf_pos}
and Ratcliff 1987).  The magnitudes at which these features occur
depend mainly on age and metallicity (and of course also on the distance
modulus in the observed LF). Both age and metallicity affect also the
SGB shape, in a way that will be examined in the following.  It is
worth noting that the turnoff is {\it not} located at the break of the
SGB, but it is $\sim 0.5$ magnitudes fainter.

In the recent years, two unexpected results raised some questions on
the ability of the models to predict the evolutionary rate out of the
MS turnoff.  In a LF obtained from the combination of
the CCD-based LFs of the three metal-poor clusters M68 (NGC~4590),
NGC~6397, and M92 (NGC~6341), Stetson (1991) found an excess of stars
on the subgiant branch just above the main-sequence turnoff, the SGB
{\it hump}, as it has been called by Faulkner \& Swenson (1993). Later
on, Bolte (1994) observed the metal-poor cluster M30 (NGC~7099) using
a mosaic of small-field CCD images and found a similar excess of SGB
stars. This excess has been investigated by several authors, both
theoretically and observationally. It has been suggested that it could
be the observable result of an unusually extended isothermal core in
turnoff stars, which could be produced by the actions of weakly
interacting massive particles (WIMPs; Faulkner \& Swenson 1993).  More
recently, different authors claimed that, at least for more metal-rich
clusters, improved observations did not show the same effect
(Sandquist et al.\ 1996; Degl'Innocenti, Weiss \& Leone 1997).

The second unexpected observational evidence involving the LFs has
been the disagreement between the theoretical and observed relative
number of MS with respect to RGB stars.  Several authors
claimed that when a theoretical LF is normalized to the MS, there is
an excess of observed giants relative to MS stars (Stetson 1991;
Bergbusch \& VandenBerg 1992; Bolte 1994; Degl'Innocenti et al.\
1997). These results have been tentatively explained by the action of
core rotation (VandenBerg, Larson, \& DePropris 1998).  It must be
noted that, at least in some cases, where the discrepancy is present
at the faint end of the MS reached by the observational data, it could
simply arise from a too steep exponent adopted for the mass
function. In a more recent paper by Silvestri et al.\ (1998), new
model LFs are presented, and a comparison of them with the observed
LFs for the clusters M5 and NGC~6397 shows that no discrepancy is
present.  The authors argue that at least part of the disagreement
claimed by previous authors could be due to a too rapid evolution of
the stars on the RGB in the models previously used, which implies a
predicted smaller number of RGB stars with respect to MS stars.

A discrepancy between observed and theoretical RGB LFs similar to 
that described above has been discussed more recently by Langer, Bolte
\& Sandquist (2000) for the two GGCs M5 and M30. They find an
overabundance of red giants in the upper RGB of these two clusters,
and they argue that the cause of this excess might be the action of
deep mixing during the stars ascent along the RGB. This phenomenon
(Sweigart 1997, 1998) should bring fresh fuel from the envelope into
the hydrogen burning shell, causing a longer RGB lifetime and
therefore an increase in the observed number of stars in that region.
However, very little evidences support the claim that the stars in M5
and M30 may have experienced a stronger deep-mixing than other
clusters that do not show similar RGB excess. In particular, a recent
paper by Rood et al.\ (1999) demonstrates that the LF of M3, another
cluster that, like M5, shows anomalous RGB abundances, interpreted as
evidences of extra-mixing (Armosky et al.\ 1994) is in perfect
agreement with the theoretical models by Straniero, Chieffi \& Limongi
(1997).

In summary, at the time we started this investigation, it was not
clear whether the discrepancies between theory and observables described 
above where real and ubiquitous, if they where confined to some
metallicity range, or if they where just an artefact due to
observational uncertainties or to the properties of the different
models used by different authors.  Because of the potential importance
of nonstandard physics in stars, and, more generally, in order to
test the evolutionary rate predicted by the models, we decided to use
the CMD database we have been creating in the last few years from our
HST GO-6095, GO-7470 and the ongoing GO-8118 snapshot survey and from
the HST archive, in order to derive accurate LFs (with large number
statistics) in a number of clusters.

There is an additional important reason why this investigation is
worth: it has been suggested (e.g., Bergbusch and VandenBerg 1992)
that the shape of the LF can be used to determine the absolute age of
GGCs. However, we will show that this application is far from being
straightforward, and the present investigation casts some doubts on
the effectiveness of the LFs to get independent and accurate distances
and ages.

For the sake of clearness, in Section 2 we will briefly summarize the
effects of the main physical parameters of a star cluster --- such as
age, metallicity and the helium content --- on the properties of the
LFs.  We will then critically discuss the most recent set of
theoretical LFs, and justify the choice of the models that will be
compared with the empirical data. In Section 3, we will present the
observed LFs. A detailed comparison with the models will follow in
Section 4. A general discussion of the implications of the present
investigation can be found in Section 5.

%_________________________ SECTION 2 _____________________________________ 

\section{The theoretical LFs}

As anticipated above, the SGB is surely the most interesting part of
the evolved-star LF, being the most sensitive to the stellar physical
parameters. In particular, the metallicity, the age and, to a minor
extent, the helium content affect both the position and the shape of
the LF. It is important to discuss in more details this dependence, if
we want to use the LF to estimate the cluster ages (see also Ratcliff
1987 and Bergbusch \& VandenBerg 1992).

\subsection{Position of the SGB break}

\begin{figure}
\centerline{\psfig{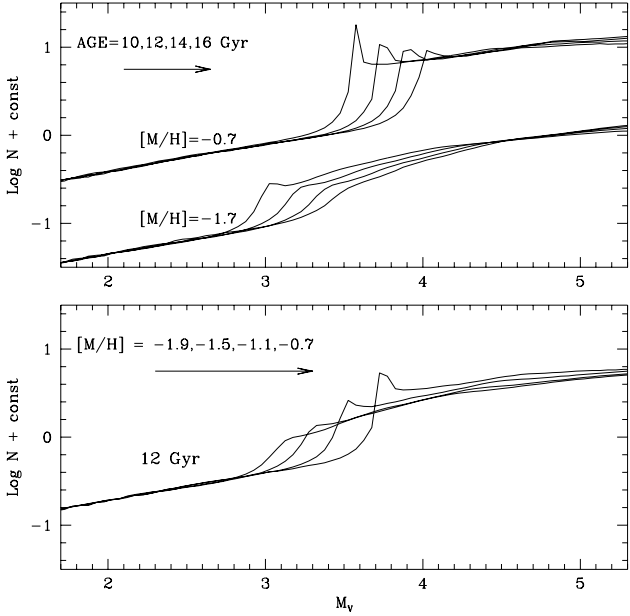}}
\caption[lf_pos.ps]{Dependence of the magnitude of the SGB break
of the LFs on the cluster age (top panel) and metallicity (bottom
panel). The model LFs are from Straniero et al. (1997) .
A vertical shift has been allowed in order to make the RGB coincide.}
\label{lf_pos}
\end{figure}

In principle, the value of using the LFs to determine or constrain
ages of GCs is due to the fact that the position of the SGB break is a
strong function of age.  Figure~\ref{lf_pos}a compares the LFs (from the
models by Straniero et al. 1997), for $t$=10,12,14,16 Gyr, for two
extremes in metallicity [M/H]=$-0.7,-1.7$, and helium abundance
$Y$=0.23. The LFs have been vertically normalized so that their RGBs
coincide. The rate at which the break moves towards lower luminosities
is essentially independent of chemical composition, but does decrease
somewhat with increasing age.  For clusters older than $\sim 12$ Gyr,
this rate is approximately 0.07 mag per Gyr, so that a determination
of the location of this feature of the LF to within $\sim 0.15$ mag
means an age uncertainty of $\sim 2$ Gyr. Unfortunately, one measures
the {\it apparent} magnitudes of the stars, so that the distance
modulus enters into the age determination as well, exactly as it does
if one tries to measure GC ages via the turnoff absolute magnitude. In
this sense, the only possible advantage of using the LF is that the
magnitude of the break is easier to determine observationally, while
the turnoff magnitude measured on a CMD is affected by an uncertainty
of the order of a few tenths of a magnitude (Rosenberg et al. 1999).
Apart from that, it must be noted that the position of the break {\it
at a given age} is a function of chemical composition, too.
Figure~\ref{lf_pos}b shows the $t=12$ Gyr LFs for $Y$=0.23 as a
function of metallicity. The values of the metallicity are indicated
in the figure label.  The dependence on metallicity is stronger than
on age, and increases with increasing metallicity.
There is also a mild dependence of the break position on the helium
content: for Y changing from 0.3 to 0.2 the break becomes $\sim 0.1$
magnitudes brighter (Ratcliff 1987).

\subsection{Shape of the SGB break}

\begin{figure}
\centerline{\psfig{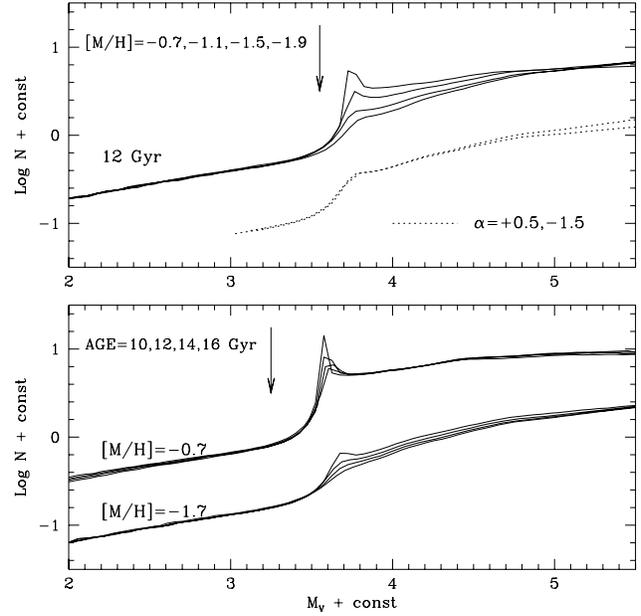}}
\caption[lf_sha.ps]{Dependence of the shape of the SGB break of the LFs
on the cluster metallicity (top panel) and age (bottom panel).
Both vertical and horizontal shifts have been allowed in order to register
the RGB and the magnitude of the break.}
\label{lf_sha}
\end{figure}

\begin{figure}
\centerline{\psfig{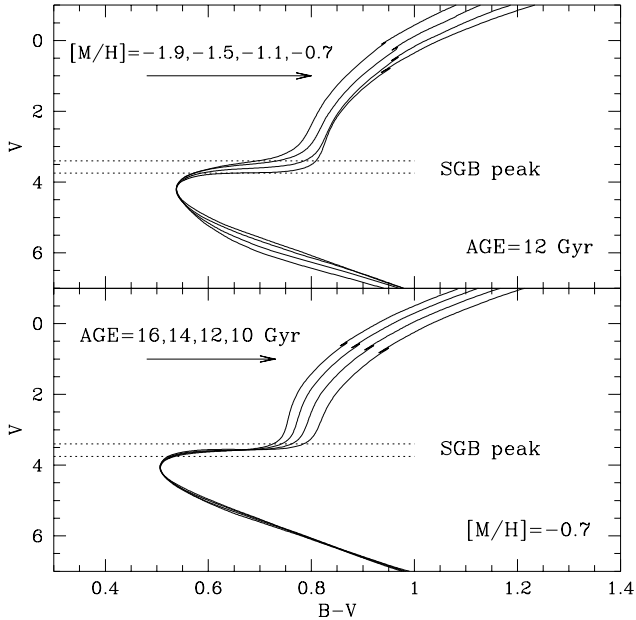}}
\caption[lf_iso.ps]{Dependence of the shape of the SGB region, in 
the CMD, as a function of cluster metallicity (top panel) and age (bottom
panel). Shifts in color, in addition to the same magnitude shift as in 
Fig.~\ref{lf_sha} have been applied in order to register the isochrones 
 at the TO magnitude.}
\label{lf_iso}
\end{figure}

The slope of the SGB region of the LF is a strong function of the
metallicity [M/H], while the shape of the SGB (i.e., the sharpness of
the peak) depends mainly on the age.  Clearly, the RGB shape is
independent of distance, making this part of the LF the most
interesting for the absolute age determination (but see the discussion
in Stetson 1991).  Figure~\ref{lf_sha} (top panel) shows how a change
in [M/H] changes the slope and the shape of the SGB and turnoff
region. Figure~\ref{lf_sha} (bottom panel) shows that the age mainly
affects the shape and the height of the peak, producing a very small
change in the slope both before and after the peak, especially for
metal-rich clusters. It must be noted, however, that beside the
observational biases that we discuss in what follows, even
from the theoretical point of view, the decrease in the sharpness 
of the SGB peak for decreasing metallicity makes the LFs of metal 
poor clusters less suitable for measuring the age from this feature.
Note that in this figure the LFs have been
registered allowing both a vertical and a horizontal shift, in order
to match the position of their RGB and SGB break. 
The cause of the changes in the LFs seen in Fig.~\ref{lf_sha} can be
better understood from Fig.~\ref{lf_iso}, which show how the SGB
region of the (theoretical) CMD is affected by a change in metallicity
at a fixed age (top panel), and by a change in age at fixed
metallicity (bottom panel). A shift in color, in addition to the same
magnitude shifts as in Fig.~\ref{lf_sha} has been applied in order to
register the isochrone at the turnoff.  The two dotted lines show the
region corresponding to the SGB peak of the LF.  It is clear from this
figure that a change in age only affects the length of the horizontal
part of the SGB, therefore the height of the SGB peak, while a change
in metallicity has a stronger effect on the shape of the isochrone
above the turnoff.
The helium content affects mainly the height of the peak. 
However the relatively small uncertainties in the GGC helium content,
combined with the small size of the effect, make the dependence on
helium of no relevance in this paper. It is also worth noting that a
change in the slope of the mass function would change the slope of the
LF only below the turnoff, i.e., $\sim 1$ mag below the SGB break
(Fig.~\ref{lf_sha} top panel, dotted lines). Therefore, different
slopes of the SGB break and of the LF, from the break to about 1
magnitude fainter, are mainly due to different chemical compositions.

\subsection{Choice of the reference models}

All the models reproduce in a similar way the main features of the LFs
discussed above. However, as one of the purposes of the present work
is to test the predicted evolutionary rates, it is of some interest to
compare the theoretical LFs published by different authors.

\begin{figure}
\centerline{\psfig{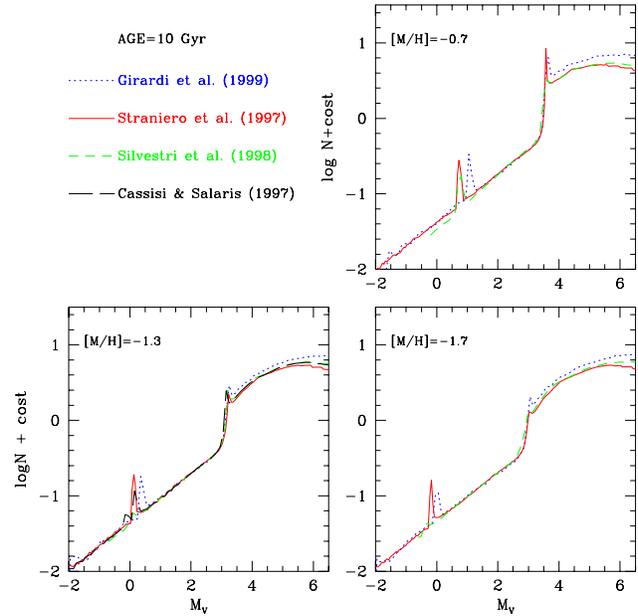}}
\caption[models.ps]{Comparison among different sets of theoretical
models. The LFs have been normalized to the region of the RGB
between the RGB bump and the SGB break by allowing a vertical
shift. All LFs refers to a mass function exponent of $\alpha=-0.5$. }
\label{models}
\end{figure}

We considered the four most recent sets of theoretical LFs covering
the whole range of metallicity spanned by our
data. Figure~\ref{models} shows the comparison of all of them, for
three typical metallicities, an age of 10 Gyr. Table~\ref{input_phys}
summarizes the main input physics of the four sets of models.  For a
detailed description of each set of models we refer the reader to the
relevant papers. As expected, the three models by Silvestri et
al. (1998), Cassisi \& Salaris (1997), and Straniero et al.\ (1997),
which have been generated with very similar input physics, are almost
indistinguishable from one another. The only difference worth noting
is that the slope of the Silvestri et al.'s models sligthly increases
towards higher metallicity (see upper right panel). This fact will be
further discussed when describing the observed LFs for the most
metal-rich clusters.  The models by Girardi et al. (1999) are slightly
different from the others, both in predicting a larger number of MS
with respect to the RGB stars, and a fainter RGB bump magnitude.
While the cause of the former is not easy to identify, the latter is
certainly due to the fact that Girardi et al. (1999) adopt
overshooting at the lower boundary of the convective envelope, and
therefore the envelope is bigger, the discontinuity in the hydrogen
profile is closer to the center, and the shell encounters it
earlier. As it will be discussed in the last section, these models
seem to better reproduce the observed RGB bump position.

\begin{table*}
\caption[]{Main model input physics}
\label{input_phys}
\begin{center}
\begin{tabular}{lllll}
\hline
Input physics   & Straniero et al.   &   Cassisi \& Salaris  &  Silvestri et al.   &
 Girardi et al.   \\
\hline
&&&&\\
%Opacity         & OPAL ($T>10^4K$)   &    OPAL ($T>10^4K$)   &   OPAL ($T>10^4K$)  &
% OPAL ($T>10^4K$)   \\
%                & AF94 ($T<10^4K$)   &    AF94 ($T<10^4K$)   &   AF94 ($T<10^4K$)  &
% AF94 ($T<10^4K$)   \\
%&&&&\\
%
He diffusion    &       yes          &          no           &          no         &
       no                \\
&&&&\\
EOS             & Straniero 1988     &   Straniero 1988      &    OPAL             &
 Straniero 1988     \\
&&&&\\
Nuclear rates   & CF88               &   CF88                &    CF88             &
   CF88          \\
                & \cdo : C85        &  \cdo : C85          &                     &
 \cdo : $1.7\times$CF88 \\
&&&&\\
Convection      &  MLT               &   MLT                 &    FST              &
 MLT             \\
                & Calib. from CSS95  &   Calib. from SC96    &    DCM97            &
 Solar calib.       \\
                &                    &                       &                     &
 + overshooting     \\
&&&&\\
Colors          & CGK97              &   CGK97               &   Kurucz 1993       &
Kurucz 1993     \\
&&&&\\
\hline
{\bf References:} &&&\\
\multicolumn{5}{l}{AF94: Alexander D.R.\ \& Fergusson J.W. 1994, ApJ, 437, 879} \\
\multicolumn{5}{l}{Straniero O. 1988, A\&AS, 76, 157} \\
\multicolumn{5}{l}{CF88: Caughlan G.R.\ \& Fowler W.A. 1988, At.\ Data Nucl.\ Data 
Tables, 40, 283} \\
\multicolumn{5}{l}{C85: Caughlan G.R., Fowler W.A., Harris, M.J., \& Zimmermann, B.A.
\ 1985, At. Data
        Nucl.\ Data Tables, 32, 197} \\
\multicolumn{5}{l}{C95: Chieffi A., Straniero O.\ \& Salaris M. 1995, ApJ, 445, L39}
\\
\multicolumn{5}{l}{SC96: Salaris M.\ \& Cassisi S. 1996, A\&A, 305, 858 }\\
\multicolumn{5}{l}{DCM97: D'Antona F., Caloi V.\ \& Mazzitelli I. 1997, ApJ, 477, 519
 }\\
\multicolumn{5}{l}{CGK97: Castelli F., Gratton R.G.\ \& Kurucz R.L. 1997, A\&A, 318,
841 }\\
\multicolumn{5}{l}{Kurucz R.L. 1993, CD-ROM 13 and CD-ROM 18}\\
\multicolumn{5}{l}{Buser R.\ \& Kurucz R.L. 1978, A\&A, 70, 555}\\
\end{tabular}
\end{center}
\end{table*}

%_________________________ SECTION 3 _____________________________________ 

\section{The data}

All the LFs presented in this paper have been obtained from our
photometric catalog, based on WFPC2 HST F439W and F555W data collected
within our GO-6095, GO-7470, and GO-8118 programs, and from the HST
archive. The CMDs for 32 of the 41 clusters observed so far can be
found in Piotto et al. (1997), Sosin et al. (1997a), and Zoccali
(1999). The details on the data reduction and calibration can be found
in Piotto et al. (1999a).  The same database has already been used for
specific investigations on the extended blue HBs (Piotto et al. 1999b),
on the RGB bump (Zoccali et al. 1999), and on the helium content from
the R parameter (Zoccali et al. 2000). All the data from the GO-6095
program published in Piotto et al.\ (1997), Sosin et al.\ (1997a), 
Sosin et al.\ (1997b), and Rich et al.\ (1997) have been reduced again
using ALLFRAME (Stetson 1994). In the original preliminary reduction
ALLSTAR (Stetson 1987) has been used. ALLFRAME allowed a slightly more
accurate photometry towards the faintest part of the CMD and to 
extend it by $\sim 1$ magnitude fainter.

Empirical LFs of the RGB+SGB+MS have been obtained for the 18 richest
clusters of our database.  Particular attention was devoted to
estimating the completeness corrections, determined by means of
artificial-star experiments. In order to optimize the use of cpu time,
we need to add the maximum number of artificial stars without creating
overcrowding, i.e., avoiding the overlap of two or more
artificial-star profiles. To this purpose, as extensively described in
Piotto \& Zoccali (1999), the artificial stars were added in a spatial
grid such that the separation of the centers in each star pair was two
PSF radii plus one pixel.  The relative position of each star was fixed
within the grid, but in each successive experiment the magnitudes of
the stars were chosen randomly and the grid was moved by a random
amount.  The input color for each artificial star was
chosen according to the fiducial points representing the CMD, while
the coordinate was transformed in order for the stars to occupy the
same position in each frame. A total of $\sim$ 4000 stars were added
(in five separate experiments) on each of the four WFPC2 chips. The
coordinate of the input and output artificial-star lists were then
transformed in order to create a single field from a mosaic of the
four WFPC2 chips.  Finally, this field was divided into three radial
annuli having approximately uniform density. The completeness
fraction was assumed to be constant inside each annulus and estimated
separately for each of them. Therefore, three LFs were obtained for
each cluster.  For the most concentrated clusters, with a strong
stellar gradient, the three LFs have different magnitude limits, and
blending effects may be present in the inner field. Only the LF points
with completeness fraction above 50\% are considered in the following.

The RGB LF was constructed as a histogram with a fixed bin size of
0.25 magnitudes. Since this region of the LF is featureless (except
for the bump, whose properties do not change radially), and very
similar in each radial annulus (due to the very small mass interval
covered, the RGB stars are not affected by mass segregation), we
summed the RGB LFs of the three annuli, in order to increase the
statistics, and to better determine both the RGB slope and the bump
location. The RGB LFs are shown as filled black dots, with errorbars,
in Figs.~5--7.  The errorbars include
the Poisson errors and the uncertainty in the completeness correction.
For the SGB and MS regions, we preferred to analyze separately the LFs
of the different radial annuli, because they are affected to different 
extents by blending and incompleteness. Furthermore, in the faintest
part of the MS, the observed LF can be affected by mass segregation,
which brings the lighter stars towards the external regions,
flattening the LF in the innermost annuli. Such effect is clearly
visible in many of our clusters, but the explored mass range is too
small (and too close to the limit magnitude) to allow a specific
analysis on the cluster dynamical properties. Of course, errors in the
completeness correction estimate could produce a similar effect on the
LF.

In order to improve the magnitude resolution, the size of the bins of
the SGB and MS LFs has been decreased with increasing number of stars.
The y-coordinate of Figs.~5--7
represents the logarithm of the number of stars per unit magnitude
interval.  In all cases, the three plotted LFs correspond, from top to
bottom, to the outer, intermediate and inner radial annulus.

\begin{table}
\caption[]{Cluster Parameters}
\label{clust_met}
\begin{center}
\begin{tabular}{cccccl}
\hline
Cluster    &  [Fe/H]     &   [M/H]   &   c   &  $\mu_{0V}$ & CMD \\
\hline
NGC 104    &  $-0.68$    &  $-0.54$  & 2.04  &  14.30  &  Sosin et al. (1997a)  \\
NGC 362    &  $-1.05$    &  $-0.84$  & 1.94  &  14.67  &  Sosin et al. (1997a)  \\
NGC1851    &  $-1.14$    &  $-0.93$  & 2.24  &  14.11  &  Sosin et al. (1997a)  \\
NGC1904    &  $-1.40$    &  $-1.19$  & 1.72  &  16.15  &  Sosin et al. (1997a)  \\
NGC2808    &  $-1.24$    &  $-1.03$  & 1.77  &  14.43  &  Sosin et al. (1997b)  \\
NGC5694    &  $-1.70$    &  $-1.49$  & 1.84  &  16.17  &  \\
NGC5824    &  $-1.69$    &  $-1.48$  & 2.45  &  14.69  &  \\
NGC5927    &  $-0.31$    &  $-0.17$  & 1.60  &  15.48  &  Sosin et al. (1997a)  \\
NGC5986    &  $-1.52$    &  $-1.31$  & 1.22  &  16.89  &  \\
NGC6093    &  $-1.48$    &  $-1.27$  & 1.95  &  14.76  &  \\
NGC6273    &  $-1.53$    &  $-1.32$  & 1.53  &  15.65  &  Piotto et al. (1999b) \\
NGC6356    &  $-0.44$    &  $-0.30$  & 1.54  &  16.06  &  \\
NGC6388    &  $-0.53$    &  $-0.39$  & 1.70  &  13.45  &  Rich et al. (1997)    \\
NGC6624    &  $-0.36$    &  $-0.22$  & pcc   &  14.45  &  Sosin et al. (1997a)  \\
NGC6652    &  $-0.86$    &  $-0.72$  & 1.80  &  15.72  &  Sosin et al. (1997a)  \\
NGC6934    &  $-1.40$    &  $-1.19$  & 1.53  &  16.99  &  Piotto et al. (1999a) \\
NGC6981    &  $-1.40$    &  $-1.19$  & 1.23  &  18.91  &  \\
NGC7078    &  $-2.03$    &  $-1.82$  & pcc   &  14.06  &  Sosin et al. (1997a)  \\
\hline
\end{tabular}
\end{center}
\end{table}

%_________________________ SECTION 4 _____________________________________ 

\section{Comparison between observational and theoretical LFs}

The LFs of each cluster are compared with a theoretical model
appropriate for its metallicity. Table~\ref{clust_met} shows the value
of [Fe/H] in the Carretta \& Gratton (1997) scale, as extended by
Cohen et al. (1999). A mean enhancement of [$\alpha$/Fe]=0.2 (for
[Fe/H]$>-1$) and [$\alpha$/Fe]=0.3 (for [Fe/H]$<-1$) was then applied
to evaluate the global metallicity [M/H] (col. 3), which has been
calculated using the relation given by Salaris, Chieffi \& Straniero
(1993). Table~\ref{clust_met} also shows the concentration parameter
(Harris 1996) and the central surface brightness in the $V$ band
(Djorgovski 1993). The last column of Table~\ref{clust_met} lists
the papers where a CMD obtained from the same data has been presented.

For all the clusters, except the four most metal-rich ones (NGC~5927,
NGC~6624, NGC~6356, and NGC~6388), we compared the empirical data only
with the models by Straniero et al. (1997). As discussed in Section~2,
a comparison with the models by Cassisi \& Salaris (1997) and by
Silvestri et al. (1998) would give identical results. As noted, the
models by Girardi et al. (1999) are slightly different from the
others, predicting a larger relative number of MS to RGB stars.
Although for most of our data the errors do not allow us to exclude
one or another theoretical prediction, some of our clusters clearly
show a better agreement with the models that predict lower MS/RGB
number ratio of stars.  The Straniero et al. models have been chosen
because they allowed the densest grid in metallicity. However, since
these models do not cover the metal-rich tail of our data, for the
four most metal-rich clusters we are forced to adopt the models by
Silvestri et al. (1998), as extended into the most metal-rich regime
by Ventura (priv. comm.).

In the comparison with the observed LFs, we adopted the following
approach: 1) For each cluster we assumed the metallicity listed in
Table~2; 2) Then we selected the age (using a grid with a 2 Gyr
separation) choosing the theoretical LF best matching the peak of the
SGB break; 3) Then we horizontally shifted the LF (i.e. found the
apparent distance modulus) in order to match the SGB break; 4) The
observed and theoretical LFs were finally vertically normalized to the
total number of stars in the featureless RGB segment between the bump
and the SGB.  

When comparing the theoretical LFs with the observed ones, we must pay
attention to the fact that there is a number of observational effects
that can modify the observed LFs. Stetson (1991) has already discussed
the possible effects of a small number statistics. However, the most
important bias in the observed LF is due to the blending, which causes
the identification of a single (brighter) star instead of two close
fainter ones. In the CMD, this phenomenon produces a spurious binary
sequence parallel to the MS, but brighter.  In the LF this translates
into a smearing of the main SGB features: the peak is lowered, the
break is less steep, and there is an overabundance of stars at the
base of the break. The latter feature corresponds to the brightest
part of the ``binary'' sequence, between the base of the RGB and the
blue stragglers (sometimes referred to as the ``yellow straggler
sequence''; Hesser et al.\ 1984). We will see that, even in the high
resolution HST images, the effects of the blending appear quite
clearly in many clusters. Typically, they are present in the inner,
more crowded region, but gradually disappear in the outer part.
As a consequence of blending we expect that {\it the age that 
we will obtain has to be considered only as an upper limit to the
cluster age}. Indeed, beside other possible systematics, like uncertainty
in the adopted metallicity scale, the only observational bias that could
artificially increase the height of the SGB peak could be an error in
the color term of the calibration equation that should distort the CMD
making the SGB more horizontal. But such an effect would also make the
slope of the observed RGB significantly steeper than what was predicted 
by the models. We do not see any such effect in our comparisons.
Finally, it is worth noting that the typical uncertainty in the 
cluster metallicity ($\pm 0.15$ dex) translates into an uncertainty
of $\sim 1.5$ Gyr in the age determination (Fig.~\ref{lf_sha}).

\begin{figure*}
\begin{tabular}{cc}
\resizebox*{0.9\columnwidth}{!}{\includegraphics{9592.f5}} &
\resizebox*{0.9\columnwidth}{!}{\includegraphics{9592.f6}} \\
\resizebox*{0.9\columnwidth}{!}{\includegraphics{9592.f7}} &
\resizebox*{0.9\columnwidth}{!}{\includegraphics{9592.f8}} \\
\resizebox*{0.9\columnwidth}{!}{\includegraphics{9592.f9}} &
\resizebox*{0.9\columnwidth}{!}{\includegraphics{9592.f10}} \\
\end{tabular}
\caption{The LFs in the three radial annuli (top to bottom: outer,
intermediate and inner) for six GGCs of our database. The radial
limits of the annuli are listed in Table~3. Overplotted
solid lines show the theoretical model for the age and metallicity
listed in the figure labels. The distance modulus required to match
the observed and theoretical SGB break position is also quoted in 
the figure label.}
\end{figure*}

\begin{figure*}
\begin{tabular}{cc}
\resizebox*{0.9\columnwidth}{!}{\includegraphics{9592.f11}} &
\resizebox*{0.9\columnwidth}{!}{\includegraphics{9592.f12}} \\
\resizebox*{0.9\columnwidth}{!}{\includegraphics{9592.f13}} &
\resizebox*{0.9\columnwidth}{!}{\includegraphics{9592.f14}} \\
\resizebox*{0.9\columnwidth}{!}{\includegraphics{9592.f15}} &
\resizebox*{0.9\columnwidth}{!}{\includegraphics{9592.f16}} \\
\end{tabular}
\caption{Same as Fig.~5, for other six GGCs of the database.}
\end{figure*}

\begin{figure*}
\begin{tabular}{cc}
\resizebox*{0.9\columnwidth}{!}{\includegraphics{9592.f17}} &
\resizebox*{0.9\columnwidth}{!}{\includegraphics{9592.f18}} \\
\resizebox*{0.9\columnwidth}{!}{\includegraphics{9592.f19}} &
\resizebox*{0.9\columnwidth}{!}{\includegraphics{9592.f20}} \\
\resizebox*{0.9\columnwidth}{!}{\includegraphics{9592.f21}} &
\resizebox*{0.9\columnwidth}{!}{\includegraphics{9592.f22}} \\
\end{tabular}
\caption{Same as Fig.~5, for other six GGCs of the database.}
\end{figure*}

In any case, the main purpose of the present investigation is not
the determination of the distance or age, though we will investigate
how the LFs can constrain these parameters. Instead, we will test the
agreement between the shape of the theoretical and observed LFs, in
order to confirm or deny the claims of disagreement mentioned in the
Introduction.

Figures~5--7 show the comparison between the observed LFs with the
best fitting theoretical ones. Note that the clusters have been
ordered in terms of metallicity, starting from the most metal rich
one. Table~3 summarizes the main results obtained for each cluster.
Column 2 shows the radial intervals where the LFs were obtained 
($r<r_1$, $r_1<r<r_2$, $r_2<r<r_3$), col.\ 3 and 4 are the metallicity 
and the age of the theoretical model. The distance modulus required to
match the SGB break and the difference between this distance and
the value quoted in the Harris (1996) catalog are listed in col.\ 5. 
Finally col.\ 6 indicates whether the observed upper RGB LFs are
reproduced by the models.

The models shown in Figs.~5-7 were all generated assuming a mass
function with slope $\alpha=-0.5$ (where Salpeter is $\alpha=-2.35$).
Such a flat mass function has been chosen because we are looking at
the cluster cores, were mass segregation causes a flattening of the
mass function. The only cluster for which a steeper mass function
($\alpha=-2.5$) was required is NGC~5694.  In almost all the GGCs for
which deep enough data are available, the effect of mass segregations
are visible in the fainter bins of the MS, the outer LF being steeper
than the inner one (see e.g.\ NGC~7078).

\begin{table*}
\caption[]{Summary of results}
\label{results}
\begin{center}
\begin{tabular}{rcccccc}
\hline
Cluster & $r_1$,$r_2$,$r_3$ & [M/H] & AGE & $(m-M)_V$ & $\Delta (m-M)_V$ &
upper \\
	&       &  (arcsec)         & (Gyr) &  (mag)  &  (mag) & RGB \\
\hline	
NGC5927    & 25,50,120 &  $-0.17$  &	12 &  15.72 &--0.09 & no   \\
NGC6624    & 15,40,120 &  $-0.30$  &	14 &  15.35 &--0.02 & no   \\
NGC6356    & 25,70,120 &  $-0.30$  &	14 &  17.08 &  0.31 & yes  \\
NGC6388    & 20,50,120 &  $-0.37$  &	14 &  16.27 &--0.27 & no   \\
NGC 104    & 20,40,120 &  $-0.60$  &	14 &  13.22 &--0.15 & yes? \\
NGC6652    & 15,50,120 &  $-0.70$  &	12 &  15.24 &  0.05 & yes  \\
NGC 362    & 30,60,120 &  $-0.80$  &	12 &  14.62 &--0.18 & yes  \\
NGC1851    & 25,50,120 &  $-0.90$  &	12 &  15.41 &--0.06 & yes? \\
NGC2808    & 25,50,120 &  $-1.00$  &	12 &  15.52 &--0.04 & yes  \\
NGC1904    & 25,60,120 &  $-1.20$  &	14 &  15.52 &--0.07 & yes  \\
NGC6934    & 18,50,120 &  $-1.20$  &	14 &  16.22 &--0.26 & yes? \\
NGC6981    & 25,60,120 &  $-1.20$  &	14 &  16.52 &  0.21 & yes  \\
NGC6093    & 15,45,120 &  $-1.30$  &	14 &  15.58 &  0.02 & yes  \\
NGC5986    & 25,50,120 &  $-1.30$  &	14 &  15.96 &  0.02 & yes  \\
NGC6273    & 25,50,120 &  $-1.30$  &	16 &  16.17 &  0.32 & yes  \\
NGC5824    & 13,45,120 &  $-1.50$  &	12 &  17.96 &  0.03 & yes  \\
NGC5694    & 13,45,120 &  $-1.50$  &	14 &  18.02 &  0.04 & yes? \\
NGC7078    & 15,40,120 &  $-1.80$  &	14 &  15.30 &--0.07 & yes  \\
\hline
\end{tabular}
\end{center}
\end{table*}

%_________________________ DISCUSSION ___________________________________

\section{Discussion} 

\subsection{The SGB hump} %_________________________________________

\begin{figure}
\centerline{\psfig{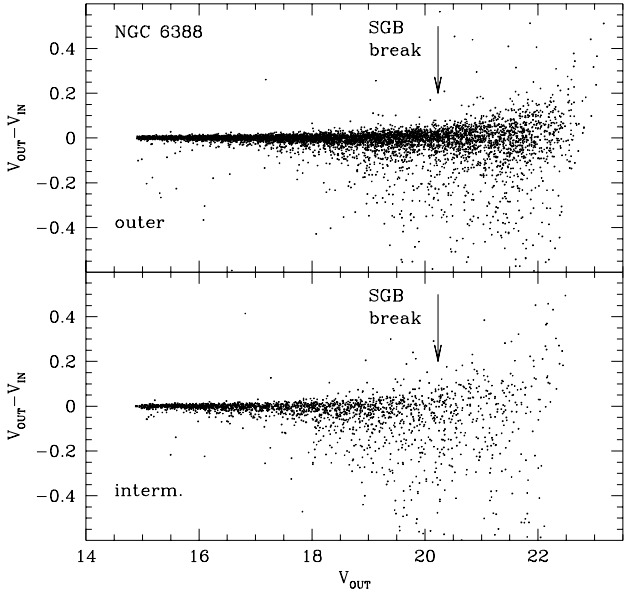}}
\caption[crowd.ps]{Difference between the input and output magnitude
of the artificial star experiments on the intermediate (lower panel) 
and outer (upper panel) annuli of NGC~6388. The vertical arrow marks
the location of the SGB break in the LF. Note that the dispersion of
the data is not symmetric, and the effect is more pronounced in the
intermediate annulus.}
\label{crowd}
\end{figure}

It is clear from Figs.~5-7 that in general there is very good
agreement between the observed and theoretical LFs. The number of
stars in the SGB, and in general the shape of this region, is very
well reproduced by the models. The only exceptions are the most
internal regions of the most concentrated clusters (e.g., NGC~1851,
NGC~2808, inner LFs) where the excess of stars just above the SGB
break is entirely due to stellar blends, that brings MS stars towards
brighter magnitudes.  This phenomenon is illustrated in
Fig.~\ref{crowd} from one of our artificial stars experiments in the
most crowded cluster NGC~6388.  In this figure, the difference between
the output and the input magnitude of the artificial stars is plotted
against the output magnitude. The stellar blends, relatively more
abundant in the intermediate annulus than in the external one, are those
stars with negative values of $V_{OUT}-V_{IN}$, i.e., measured with a
magnitude brighter than the original one. The vertical arrow marks the
magnitude of the SGB break. At this magnitude level, in the
intermediate annulus, about half of the stars are found in blends,
while in the outer annulus this fraction is considerably
reduced. Fig.~5 confirms that the LF for NGC~6388 is of poor
quality, especially in the innermost regions.  The cause of this is
partially due to the presence of both total and differential reddening
that broadens the CMD sequences (see Rich et al.\ 1997), but the excess
of stars above the SGB break is clearly due to crowding, since it
gradually disappears towards the outer part.

We conclude that our data do not support the claims of the 
presence of a SGB hump. We believe that the most plausible explanation 
for this effect is the presence of stellar blends, strongly affecting
most ground-based studies.

\subsection{The ratio of MS/RGB stars} %____________________________

The observed number of MS/RGB stars is in agreement with theoretical
predictions, within the uncertainties. As concluded by Degl'Innocenti 
et al. (1997), minor deviations in some cases (e.g.\ NGC~6981, outer LF)
are more likely the consequence of mass segregation, or of residual errors 
in the completeness corrections or simply to small number statistics along 
the RGB were the LFs were normalized, than real features to ascribe to a 
mismatch with theory. Indeed, such discrepancies disappear in the
most internal, richest LFs. Of course we cannot exclude the possibility
of a real radial gradient in the number ratio of MS/RGB stars, but
this hypothesis does not seem very plausible and would require further tests
with wider radial coverage.

\subsection{The upper RGB} %___________________________________

The problem of the upper RGB is more complicated. At variance with what
was found by Langer et al.\ (2000), we do not find any systematic excess
of RGB stars for [M/H]$>-0.7$. Some excess of RGB stars similar to
that found by Langer et al.\ (2000) is visible for the clusters
NGC~6652, NGC~1851, NGC~6934, and NGC~5694, but in {\it all} these
cases the theoretical models agree with the data within the errors, so
these discrepancies are not significant.  In the more metal rich
regime ([M/H]$<0.7$), and in particular for NGC~5927, NGC~6624 and
NGC~6388, we find systematic deviations of different nature. While
Langer et al.\ (2000) find a displacement of the observed LF with
respect to the theoretical one, for these three clusters we find that
the observed upper RGB slopes are flatter than the theoretical ones.
It must be noted that in this interval we used a different set of
models (Silvestri et al. 1998), which become steeper and steeper in
the upper RGB with increasing metallicity (cf. Fig.~\ref{models}).  It
would be interesting to verify whether also the other models show the
same behaviour in the most metal-rich regime. In any case, this slope
change is much smaller than the observed discrepancy between data and
theory (Fig.~5). Possible alternative explanations are: 1)
contamination by AGB stars, 2) the presence of differential reddening,
and 3) problems with the bolometric corrections adopted in the
models. We exclude the first two possibilities. Indeed, the expected
number of AGB stars (most of which have been excluded from the LF)
cannot explain a difference by a factor of two, as observed in the
upper bins of the most metal rich clusters of Fig.~5.  We have also
performed numerical experiments to test the second possibility, and
found that the presence of differential reddening causes a smoothing
of the LF main features but not a change in its slope. Instead, the
fact that the deviation between observations and theory is confined to
a gradually smaller and brighter magnitude interval for decreasing
metallicity, suggests that this mismatch could be due to problems in
the bolometric correction. Indeed, as shown in Ortolani et al. (1990,
1992), the upper RGB for the most metal rich clusters in the $V$
vs. ($B-V$) plane turns down due to blanketing effects. As already
discussed by Ortolani et al. (1990), and verified also for the models
adopted here, even the most recent bolometric corrections are not able
to correctly reproduce this feature in the CMD. For the same reason,
we expect that also the theoretical LFs for the most metal rich
clusters cannot match the observed data in brightest bins.  In view
of the potential implications of this discrepancy in terms of the
evolutionary lifetime in the upper RGB of metal-rich clusters, this
effect deserves further investigations.

\subsection{Distances and Ages} %____________________________________

\begin{figure}
\centerline{\psfig{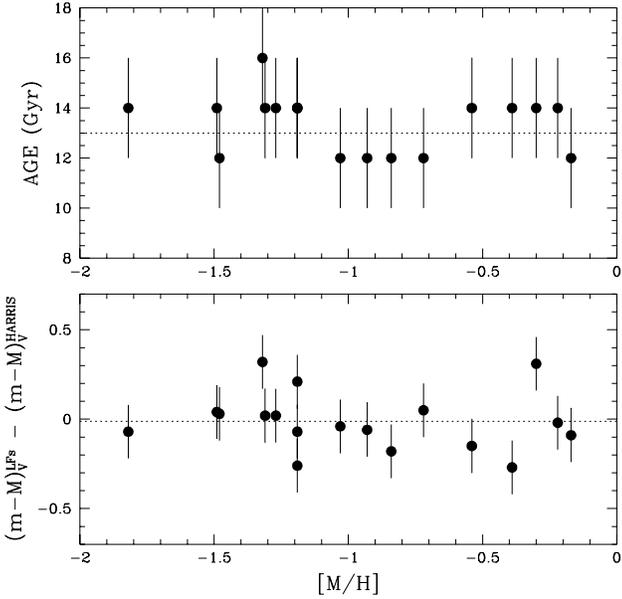}}
\caption[age_dist.ps]{{\it Upper panel}: the measured cluster ages
as a function of metallicity. {\it Lower Panel}: the difference between 
the distance moduli of the studied clusters given in Harris's catalogue, 
and the distance moduli determined from the comparison between the 
observed and the theoretical LFs.}
\label{age_dist}
\end{figure}

The approach we followed in the comparison between the observed and
theoretical LFs has been to {\it measure} the age from the shape of
the SGB peak, and then to derive the distance modulus required to
match the SGB break at that fixed age. As we discussed in Section~2
this is not the best method to measure cluster ages and distances,
since it is strongly affected by the quality of the photometric
data. Still, it might be of some interest to compare the distances
obtained from our LFs with the ones in the literature.  In the {\it
lower panel} of Fig.~\ref{age_dist}, we compare our distance moduli
with the values in the most updated Harris's catalog (June 22$^{nd}$,
1999 revision; 

http://physun.physics.mcmaster.ca/GC//mwgc.dat). It is
reassuring that there is a general agreement between the two sets of
distances.  The average difference is: $(m-M)_V^{\rm LFs}-(m-M)_V^{\rm
HARRIS}=-0.01\pm 0.15$, and the residuals show no trend with the
metallicity.  The logic consequence of this agreement is that also the
ages should be on average correct.  The {\it upper panel} of
Fig.~\ref{age_dist} shows the ages found for each cluster with the
method described above. The average age is 13 Gyr, while the
dispersion is of the same order of the error bars, and there is no age
trend with the metallicity.  Keeping in mind the large uncertainty
related to our age determination, it is of some interest to note that
we have the same result on GGC relative ages as in Rosenberg et
al. (1999).

\subsection{The RGB bump} %____________________________________________

Another feature of the observed LF that must be compared with
theoretical models is the RGB bump.  There have been claims that the
observed position of the RGB bump is not well reproduced by
theoretical models. The first time that this feature was clearly
identified in a GGC (King et al.\ 1985), its position was found to be
0.85 magnitudes fainter than predicted by the models of Sweigart \&
Gross (1978). Later on, in a dedicated paper, Fusi Pecci et al.\
(1990) found a smaller average discrepancy ($\sim 0.42$ magnitudes)
from the comparison of a large sample of data (11 GGCs) with the
models by Rood \& Crocker (1989).  More recently, Cassisi \& Salaris
(1996) and Zoccali et al.\ (1999) showed that, adopting the new
metallicity scale by Carretta \& Gratton (1997) and taking into
account a modest alpha-element enhancement, the observed magnitude
difference between the bump and the HB (\vhbb) is in good agreement
with canonical theoretical models that include the most updated input
physics.

Although it is true that the \vhbb is well reproduced by the models,
Figs.~5--7 clearly shows that the absolute magnitude of the
theoretical RGB bump is still systematically brighter than the
observed one by $\sim 0.25$ mags.  Of course, the level of agreement
between the theoretical and empirical bump location strongly depends
on the adopted distance modulus. We already showed that our
derived distance moduli are on average in good agreeement with the
values obtained with the distance scale adopted in Harris's catalog
(Fig.~\ref{age_dist}).

Only the models by Girardi et al.\ (1999) are able to correctly locate
the bump. However, it is worth noting that this result has been obtained 
by including in the models an additional free parameter, i.e., some
overshooting at the base of the convective envelope (Bressan, Chiosi, 
\& Bertelli 1981).

\begin{figure}
\centerline{\psfig{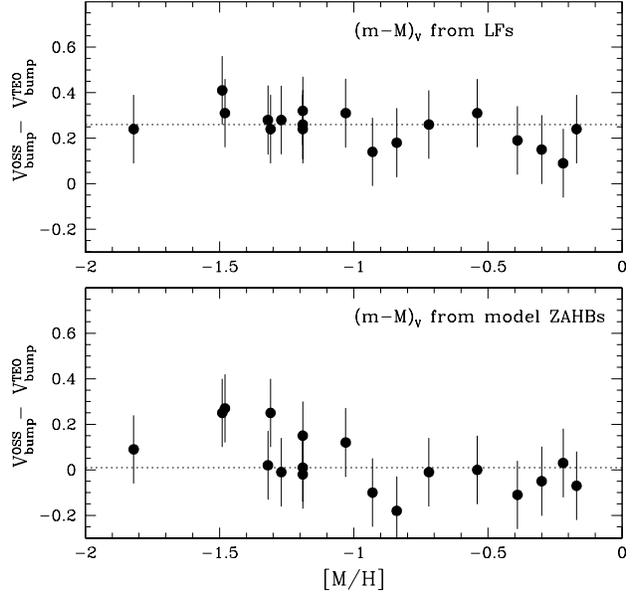}}
\caption[delta.bump.ps]{Difference between the observed and
theoretical location of the RGB bump, as a function of metallicity,
for two different assumption on the distance. Top panel: the
adopted distance moduli have been determined via the comparison
between theoretical and observed LFs. Bottom panel: the distance
moduli come from the comparison between observed and theoretical
ZAHBs.}
\label{delta.bump}
\end{figure}

Figure~\ref{delta.bump} ({\it upper panel}) shows that the theoretical
magnitude of the bump as predicted by the LFs of Straniero et al.\
(1997) and Silvestri et al.\ (1998) is systematically brighter by 0.26
mags.  On the other hand, the fact that the observed \vhbb is in good 
agreement with the theoretical predictions suggests that adopting the
distance modulus obtained from the theoretical absolute magnitude 
of the zero age horizontal branch (ZAHB), we are able to remove this 
discrepancy.

Indeed, from a comparison between the theoretical ZAHB location (Cassisi 
et al.\ 1999):
\[
M_V^{ZAHB} = 0.974 + 0.379 \times {\rm [M/H]} + 0.062 \times {\rm
[M/H]}^2
\]
and the empirical ZAHB magnitudes of Zoccali et al. (1999), we would
obtain systematically longer distances (by $\sim 0.26$ mag) and
therefore a very good match of the position of the RGB bump
(Fig.~\ref{delta.bump}, {\it lower panel}).  It must be noted that the
distances are also systematically longer than the values in the
Harris catalog\footnote{The distances in the latest issue of the
Harris catalog are computed from the observed mean magnitude of the HB, 
adopting the relation $M_V(HB)=0.15\mbox{[Fe/H]}+0.80$.}, 
but more consistent with the HIPPARCOS distance scale (Salaris \& Weiss 1998).

Of course, adopting the longer distance scale implies a $\sim 2-4$ Gyr
younger age for all the clusters. Then one is left with the impression
that, if the SGB shape and the ZAHB give different results for the age
and the distance, there must be some internal inconcistency in the
theoretical models.  However, such a conclusion is premature,
given that the ages measured via the SGB shape must be considered only
an upper limit, therefore the LF approach do not allow us to exclude
younger ages and longer distances, as required to fit the ZAHB models
(and the RGB bump).

%_________________________ CONCLUSIONS _______________________________________ 

\section{Conclusions}

We obtained the LFs for the RGB+SGB+MS of a sample of 18 GGCs, observed 
with HST, with the main purpose of verifying whether the claimed LF 
anomalies (Stetson 1991, Faulkner and Swenson 1993, Bolte 1994 and Langer
et al.\ 2000) were real or not. We also attempted a determination of
the cluster ages from the shape of the SGB peak, and from this we
inferred a value for the cluster distance. Our main conclusions are
the following:

\begin{itemize}
\item
No SGB hump is found in any of our GGCs. We showed that a very similar
feature is present in the most crowded regions of our clusters, and 
therefore we conclude that it has to be interpreted as an artifact of
stellar blending;
\item
The observed number of MS/RGB stars is on average in good agreement 
with theoretical predictions in all the explored metallicity range;
\item
There is no statistically significant excess of stars in the observed 
upper RGB LF, at any metallicity;
\item
A difference in the slope of the upper RGB LF of the most metal rich
clusters ([M/H]$>-0.7$) is likely to be due to errors in the
bolometric corrections for extremely red stars. Indeed, also the CMD
of such metal rich clusters do not agree with theoretical models in
the upper RGB region, for the same reason;
\item
Although in principle the shape of the SGB allows the determination
of the cluster age, in practice the method is strongly affected both
by errors in the metallicity, and, even more, by the quality of the
photometry. Keeping in mind this uncertainty, we find a mean age of
13 Gyr, without any trend with metallicity.
\item
The cluster distances we obtain after imposing the ages derived above
are on average in good agreement with the distances quoted in the
latest revision of the Harris catalog, the difference being
$(m-M)_V^{\rm LFs}-(m-M)_V^{\rm HARRIS}=-0.01\pm 0.15$.
\item
We find a systematic difference of $\sim 0.26$ magnitudes between the
empirical and theoretical absolute magnitude of the RGB bump, with the
exception of the models by Girardi et al.\ (1999). The discrepancy
disappear if we adopt the longer distance scale obtained from
theoretical ZAHB models (Cassisi et al.\ 1999).
\end{itemize}

%_________________________ ACKNOWLEDGEMENTS __________________________________ 

\begin{acknowledgements}

We want to thank the referee, Alvio Renzini, for his helpful comments
and discussions, which surely improved in many aspects this paper.
We are indebted with Alessandro Chieffi, for having provided the
programs to interpolate within the grid of the Straniero et al's
models, and to Paolo Ventura, who specifically calculated the models
for the most metal-rich clusters. We also thank Leo Girardi and Peter
Stetson for many useful discussions. It is a pleasure to thank Santino
Cassisi for being constantly available for discussions and suggestions,
and Giuseppe Bono for useful comments on the original manuscript. 
This work was supported by MURST under the project:
"Stellar Dynamics and Stellar Evolution in Globular Clusters".
Partial support by ASI is also acknowledged.

\end{acknowledgements}

%_________________________ REFERENCES __________________________________ 


\begin{thebibliography}{}
\bibitem{} Armosky, B.~J., Sneden, C., Langer, G.~E., Kraft, R.~P., 1994,
	AJ, 108, 1364
\bibitem{} Bergbusch, P., \& VandenBerg, D.\ A., 1992, ApJS, 81, 163
\bibitem{} Bolte, M. 1994, ApJ, 431, 223
\bibitem{} Bressan, A.~G., Chiosi, C., \& Bertelli, G.\ 1981, A\&A, 102, 25
\bibitem{} Carretta, E., \& Gratton, R.\ G.\ 1997, A\&AS, 121, 95
\bibitem{} Cohen, J.\ G., Gratton, R.\ G., Behr, B.\ B., Carretta, 
	E.\ (1999), ApJ, 523, 739
\bibitem{} Degl'Innocenti, S., Weiss, A. \& Leone, L.\ 1997, A\&A, 319, 
	487
\bibitem{} Djorgovski, S.\ G.\ 1993, in ASP Conf.Ser.50, Structure and 
	Dynamics of Globular Clusters, ed. Djorgovski, S., Meylan, G. 
	(San Francisco: ASP), 373
\bibitem{} Faulkner, J. \& Swenson, F.\ J.\ 1993, ApJ, 411, 200
\bibitem{} Fusi Pecci, F., Ferraro, F.~R., Crocker, D.~A., Rood, R.~T.,
	\& Buonanno, R. 1990, A\&A, 238, 95
\bibitem{} Girardi, L., Bressan, A., Bertelli, G., Chiosi, C.\ 1999, 
	A\&A, in press (astro-ph/9910164)
\bibitem{} Gratton, R.\ G., Fusi Pecci, F., Carretta, E., Clementini, 
	G., Corsi, C.\ E., Lattanzi, M.\ 1997, ApJ, 491, 749
\bibitem{} Harris, W.\ E. 1996, AJ, 112, 1487
\bibitem{} Hesser, J.~E., McClure, R.~D., Hawarden, T.~G., Cannon, R.~D,
	von Rudloff, R., Kruger, B., \& Egles, D. 1984, P.~A.~S.~P., 96, 
	406
\bibitem{} King, C.~R., Da Costa, G.~S., \& Demarque, P., 1985, ApJ, 299,
	674
\bibitem{} Langer, G.~E., Bolte, M., \& Sanquist, E., 2000, ApJ, 529, 936
\bibitem{} Ortolani, S., Barbuy, B., \& Bica, E. 1990, A\&A, 236, 362
\bibitem{} Ortolani, S., Bica, E., \& Barbuy, B. 1992, A\&AS, 92, 441 
\bibitem{} Piotto G., et al.\ 1997, in Advances in Stellar Evolution,
	eds. R.\ T.\ Rood \& A.\ Renzini (Cambridge: Cambridge 
	University Press), p.\ 84.
\bibitem{} Piotto, G. et al. 1999a, AJ, 117, 264
\bibitem{} Piotto, G. et al. 1999b, AJ, 118, 1737
\bibitem{} Piotto, G., \& Zoccali, M.\ 1999, A\&A, 345, 485
\bibitem{} Ratcliff S.\ J. 1987, ApJ, 318, 196
\bibitem{} Reid, I.\ N. 1998, AJ, 115, 204
\bibitem{} Renzini, A., \& Fusi Pecci, F.\ 1988, ARA\&A, 26, 199
\bibitem{} Rich, R.\ M., Sosin, C., Djorgovski, S.\ G., Piotto, G., King,
	I.\ R., Renzini, A., Phinney, E.\ S., Dorman, B., Liebert, J., \& 
	Meylan, G. 1997, ApJL, 484, L25 
\bibitem{} Rood R.T. \& Crocker D.A., 1989, in The Use of Pulsating Stars
	in Fundamental Problems of Astronomy, IAU Coll.\ 111, ed. E.G.
	Schmidt, Cambridge U. Press, Cambridge, UK, p. 103
\bibitem{} Rood, R.~T., Carretta, E., Paltrinieri, B., Ferraro, F.~R.,
	Fusi Pecci, F., Dorman, B., Chieffi, A., Straniero, O., \&
	Buonanno, R., 1999, ApJ, 523, 752 
\bibitem{} Salaris, M., Chieffi, A., \& Straniero, O. 1993, ApJ, 414, 580
\bibitem{} Salaris, M., \& Weiss, A., 1998, A\&A, 335, 943
\bibitem{} Sandquist, E.\ L., Bolte, M., Stetson, P.\ B., Hesser, J.\ E.\ 
	1996, ApJ, 470, 910
\bibitem{} Silvestri, F., Ventura, P., D'Antona, F. \& Mazzitelli, I. 1998, 
	ApJ, 509, 192
\bibitem{} Sosin, C., Piotto, G., King, I.~R., Djorgovski, S.~G.,
        Rich, R.~M., King, I.~R., Dorman, B., Liebert, J., \& Renzini,
        A. 1997a, in Advances in Stellar Evolution, eds.\ R.\ T.\ Rood
        and A.\ Renzini (Cambridge:\ Cambridge Univ.\ Press), p. 92
\bibitem{} Sosin, C., Dorman, B., Djorgovski, S.~G., Piotto, G., Rich,
        R.~M., King, I.~R., Liebert, J., Phinney, E.~S., \& Renzini, A.
        1997b, ApJL, 480, L35
\bibitem{} Stetson, P.\ B., in ASP Conf.Ser.13, The Formation and Evolution 
	of Star Clusters, ed.\ K.\ Janes (San Francisco: ASP), 88
\bibitem{} Straniero, O., Chieffi, A., \& Limongi, M. 1997, ApJ, 490,
	425
\bibitem{} Sweigart, A.~V., \& Gross, P.~G., 1978, ApJS, 36, 405
\bibitem{} Rosenberg, A.~R., Saviane, I., Piotto, G., Aparicio, A.\ 1999, 
	AJ, 118, 2306
\bibitem{} VandenBerg, D.\ A., Larson, A.\ M., \& DePropris, R.\ 1998, PASP, 
	110, 98
\bibitem{} Zoccali, M.\ 1999, Tesi di Dottorato di Ricerca, Universit\`a di 
	Padova
\bibitem{} Zoccali, M., Cassisi, S., Piotto, G., Bono, G., Salaris, M.\ 1999,
	ApJ, 518, L49
\bibitem{} Zoccali, M., Cassisi, S., Bono, G., Piotto, G., Rich, R.~M., \& 
	Djorgovski, S.~G., 2000, ApJ, in press

\end{thebibliography}
\end{document}